\documentclass[12pt]{article}
\usepackage{fancyhdr}

\usepackage{mathrsfs}
\usepackage[T1]{fontenc}
\usepackage{mathpazo}
\usepackage{setspace}
\usepackage{amsfonts}
\usepackage{amssymb}
\usepackage{amsmath}
\usepackage{epsfig}
\usepackage{latexsym}
\usepackage{color}
\usepackage{graphicx}
\usepackage{nicefrac}
\usepackage[latin1]{inputenc}
\usepackage{slashed}
\usepackage{multirow}
\usepackage{comment}
\usepackage{soul}
\usepackage{hyperref}
\usepackage[nosort]{cite}
\usepackage{datetime}
\usepackage{braket}

\usepackage[titletoc]{appendix}
\def\hybrid{\topmargin -20pt    \oddsidemargin 0pt
        \headheight 0pt \headsep 0pt
        \textwidth 6.25in       % A4 paper
        \textheight 9.25in       % A4 paper
        \marginparwidth .875in
        \parskip 5pt plus 1pt   \jot = 1.5ex}

\hybrid

\def\baselinestretch{1.2}

\catcode`\@=11

\def\marginnote#1{}
\newcount\hour
\newcount\minute
\newtoks\amorpm
\hour=\time\divide\hour by60
\minute=\time{\multiply\hour by60 \global\advance\minute by-\hour}
\edef\standardtime{{\ifnum\hour<12 \global\amorpm={am}%
        \else\global\amorpm={pm}\advance\hour by-12 \fi
        \ifnum\hour=0 \hour=12 \fi
        \number\hour:\ifnum\minute<10 0\fi\number\minute\the\amorpm}}
\edef\militarytime{\number\hour:\ifnum\minute<10 0\fi\number\minute}
\def\draftlabel#1{{\@bsphack\if@filesw {\let\thepage\relax
   \xdef\@gtempa{\write\@auxout{\string
      \newlabel{#1}{{\@currentlabel}{\thepage}}}}}\@gtempa
   \if@nobreak \ifvmode\nobreak\fi\fi\fi\@esphack}
        \gdef\@eqnlabel{#1}}
\def\@eqnlabel{}
\def\@vacuum{}
\def\draftmarginnote#1{\marginpar{\raggedright\scriptsize\tt#1}}

\def\draft{\oddsidemargin -.5truein
        \def\@oddfoot{\sl preliminary draft \hfil
        \rm\thepage\hfil\sl\today\quad\militarytime}
        \let\@evenfoot\@oddfoot \overfullrule 3pt
        \let\label=\draftlabel
        \let\marginnote=\draftmarginnote
   \def\@eqnnum{(\theequation)\rlap{\kern\marginparsep\tt\@eqnlabel}%
\global\let\@eqnlabel\@vacuum}  }

\def\preprint{\twocolumn\sloppy\flushbottom\parindent 2em
        \leftmargini 2em\leftmarginv .5em\leftmarginvi .5em
        \oddsidemargin -.5in    \evensidemargin -.5in
        \columnsep .4in \footheight 0pt
        \textwidth 10.in        \topmargin  -.4in
        \headheight 12pt \topskip .4in
        \textheight 6.9in \footskip 0pt
        \def\@oddhead{\thepage\hfil\addtocounter{page}{1}\thepage}
        \let\@evenhead\@oddhead \def\@oddfoot{} \def\@evenfoot{} }

\def\numberbysection{\@addtoreset{equation}{section}
        \def\theequation{\thesection.\arabic{equation}}}

\def\underline#1{\relax\ifmmode\@@underline#1\else
        $\@@underline{\hbox{#1}}$\relax\fi}

\def\titlepage{\@restonecolfalse\if@twocolumn\@restonecoltrue\onecolumn
     \else \newpage \fi \thispagestyle{empty}\c@page\z@
        \def\thefootnote{\fnsymbol{footnote}} }

\def\endtitlepage{\if@restonecol\twocolumn \else \newpage \fi
        \def\thefootnote{\arabic{footnote}}
        \setcounter{footnote}{0}}  %\c@footnote\z@ }

\catcode`@=12
\relax

\def\figcap{\section*{Figure Captions\markboth
        {FIGURECAPTIONS}{FIGURECAPTIONS}}\list
        {Figure \arabic{enumi}:\hfill}{\settowidth\labelwidth{Figure
999:}
        \leftmargin\labelwidth
        \advance\leftmargin\labelsep\usecounter{enumi}}}
 \relax
\def\tablecap{\section*{Table Captions\markboth
        {TABLECAPTIONS}{TABLECAPTIONS}}\list
        {Table \arabic{enumi}:\hfill}{\settowidth\labelwidth{Table
999:}
        \leftmargin\labelwidth
        \advance\leftmargin\labelsep\usecounter{enumi}}}
 \relax
\def\reflist{\section*{References\markboth
        {REFLIST}{REFLIST}}\list
        {[\arabic{enumi}]\hfill}{\settowidth\labelwidth{[999]}
        \leftmargin\labelwidth
        \advance\leftmargin\labelsep\usecounter{enumi}}}
 \relax
\makeatletter
\newcounter{pubctr}
\def\publist{\@ifnextchar[{\@publist}{\@@publist}}
\def\@publist[#1]{\list
        {[\arabic{pubctr}]\hfill}{\settowidth\labelwidth{[999]}
        \leftmargin\labelwidth
        \advance\leftmargin\labelsep
        \@nmbrlisttrue\def\@listctr{pubctr}
        \setcounter{pubctr}{#1}\addtocounter{pubctr}{-1}}}
\def\@@publist{\list
        {[\arabic{pubctr}]\hfill}{\settowidth\labelwidth{[999]}
        \leftmargin\labelwidth
        \advance\leftmargin\labelsep
        \@nmbrlisttrue\def\@listctr{pubctr}}}
 \relax
\makeatother
\newskip\humongous \humongous=0pt plus 1000pt minus 1000pt

\newif\ifdtup

\relax

\def\be{\begin{equation}}
\def\ee{\end{equation}}
\def\ba{\begin{eqnarray}}
\def\ea{\end{eqnarray}}

\def\del{\partial}

\def\b{\beta}

\def\g{\gamma}
\def\G{\Gamma}
\def\d{\delta}

\def\m{\mu}

\def\om{\omega}

\def\l{\lambda}

\def\s{\sigma}

\def\cL{{\cal L}}

\def\no{\noindent}

\def\qq{\qquad}

\def\IR{\relax{\rm I\kern-.18em R}}

\def \z { {\bar z} }

\def \J {{\bar J} }

\def \ha {{1\over 2}}

\def \ov {\over}

\def\diag{{\rm diag}}

\def\IR{\relax{\rm I\kern-.18em R}}
\def\IL{\relax{\rm I\kern-.18em L}}

\def\inv{^{\raise.15ex\hbox{${\scriptscriptstyle -}$}\kern-.05em 1}}

\def\cL{{\cal L}}

\begin{document}
%Text\fontsize{13}{12}\selectfont Text

%\renewcommand{\theequation}{\arabic{equation}}
\renewcommand{\theequation}{\thesection.\arabic{equation}}
\csname @addtoreset\endcsname{equation}{section}

\newcommand{\beq}{\begin{equation}}
\newcommand{\eeq}[1]{\label{#1}\end{equation}}
\newcommand{\ber}{\begin{equation}}
\newcommand{\eer}[1]{\label{#1}\end{equation}}
\newcommand{\eqn}[1]{(\ref{#1})}
\begin{titlepage}
\begin{center}

\hfill CERN-TH-2018-177
%\vskip -.1 cm
%\hfill hep--th/yymmnnn\\

\vskip .3 in

{\large\bf Weyl anomaly and the $C$-function in $\lambda$-deformed CFTs}

\vskip 0.35in

{\bf Eftychia Sagkrioti}$^1$, {\bf Konstantinos Sfetsos}$^{1}$, {\bf Konstantinos Siampos}$^2$
\vskip 0.1in

{\em${}^1$Department of Nuclear and Particle Physics,\\ Faculty of Physics, National and Kapodistrian University of Athens,\\15784 Athens, Greece
}

\vskip 0.1in

{\em${}^2$Theoretical Physics Department, CERN, 1211 Geneva 23, Switzerland
}

\vskip 0.1in

{\footnotesize \texttt esagkrioti@phys.uoa.gr, ksfetsos@phys.uoa.gr, konstantinos.siampos@cern.ch}

\vskip .43in
\end{center}

\centerline{\bf Abstract}
\noindent
For a general $\lambda$-deformation of current algebra CFTs we compute the exact Weyl anomaly coefficient and the corresponding metric in the couplings space geometry. By incorporating the exact $\beta$-function found in previous works
we show that the Weyl anomaly is in fact the exact Zamolodchikov's $C$-function interpolating between exact CFTs occurring in the UV and in the IR. We provide explicit examples with the anisotropic $SU(2)$ case presented in detail.
The anomalous dimension of the operator driving the deformation is also computed in general.
Agreement is found with special cases existing already in the literature.

\no

\vskip .4in
\noindent
\end{titlepage}
\vfill
\eject

\newpage
%\vskip .3in

\tableofcontents

\noindent

\def\baselinestretch{1.2}
\baselineskip 20 pt
\noindent

%%%%%%%%%%%%%%%%%%%%%%%%%%%%%%%%%%%%%%%%%%

\setcounter{equation}{0}
\renewcommand{\theequation}{\thesection.\arabic{equation}}

\section{Introduction}
\label{secintro}

According to Zamolodchikov's $c$-theorem \cite{Zamolo:1986}, for a two-dimensional renormalizable quantum field theory (QFT) there is a positive function of the couplings, called the $C$-function, which monotonically decreases
under the renormalization group (RG) flow of the theory from the UV to the IR. At the fixed points of the RG flow { the $C$-function} equals the central charges of the corresponding conformal field theories (CFTs). Since  the stress-energy tensor couples to all
degrees of freedom of a theory, the $C$-function is associated with the degrees of freedom of the theory 
at a certain energy scale. Thus, the physical interpretation of
the $c$-theorem is that by flowing towards lower energy scale, progressively more information is lost.
Due to the fact that the degrees of freedom are integrated out during the flow, this information loss is
irreversible. More intuitively, as the energy scale decreases,  heavier degrees of freedom decouple from the low-energy dynamics of the theory, hence leading to a monotonically decreasing $C$-function.

In a generic QFT with couplings $\lambda_i$, the $C$-function obeys \cite{Zamolo:1986}
\begin{equation}
\label{Cfunctiongeneral}
\frac{\text{d}C}{\text{d}t}=\b^i\partial_iC=24G_{ij}\b^i\b^j\geqslant0\,,\quad \b^i=\frac{\text{d}\lambda_i}{\text{d}t}\,,\quad t=\text{ln}\m^2\ ,
\end{equation}
where $G_{ij}$ is the Zamolodchikov metric in the space of couplings.
For convenience we have used subscript indices in the $\l$'s in order to simplify the expressions and to follow convention used in
literature.

Recently, the first examples in literature where the $C$-function has been computed exactly as a function of the couplings were
found \cite{c-function:2018}.
Specifically, this involved the $\l$-deformed models
of \cite{Sfetsos:2013wia}  and \cite{Georgiou:2016urf,Georgiou:2017jfi} based on one or two WZW models, respectively, for the special isotropic cases having one or two deformation parameters. In this research line the main goals we achieve with the present paper are: As a follow-up to \cite{c-function:2018}, we present the exact $C$-function for the aforementioned  (doubly) $\l$-deformed models but for generic couplings. In addition, we compute the metric
in the couplings space geometry, which has potential uses beyond the present paper, as well as the anomalous dimension matrix of the composite operator driving the perturbation away from the conformal point. In the process
we show that the $C$-function is in fact the Weyl anomaly coefficient computed by cleverly utilizing $\s$-model data corresponding to the $\l$-deformations.

The action of the doubly deformed models \cite{Georgiou:2017jfi} represents the effective action of two WZW models at different Kac--Moody levels $k_1$ and $k_2$, mutually interacting via current bilinears
\begin{equation}
S_{k_1,k_2}^{\lambda_1,\lambda_2}=S_{k_1}(\mathfrak{g}_1)+S_{k_2}(\mathfrak{g}_2)+\frac{k}{\pi}\int \text{d}^2\sigma\,
\big((\lambda_1)_{ab}J_{1+}^aJ_{2-}^b+(\lambda_2)_{ab}J_{2+}^aJ_{1-}^b\big)+\cdots\ ,
\label{action}
\end{equation}
  where $k=\sqrt{k_1k_2}$ and $S_{k_i}$, $i=1,2$, are the WZW actions for a group elements $\mathfrak{g_i}\in G$, of a semi-simple, compact and simply connected Lie group $G$.
 The currents are
  \begin{equation}
  J^a_{i+}=-i\,\text{Tr}(t^a\partial_+\mathfrak{g}_i\mathfrak{g}_i^{-1})\,,\quad
  J^a_{i-}=-i\,\text{Tr}(t^a\mathfrak{g}_i^{-1}\partial_-\mathfrak{g}_i)\,,\quad i=1,2\,.
  \end{equation}
The $t^a$'s are Hermitian matrices{ ,  normalized
as $\text{Tr}\left(t^at^b\right)=\d_{ab}$ and they obey 
$[t^a,t^b]=if_{abc}t^c$, where the structure constants $f_{abc}$'s are taken to be real.}

The effective action incorporating all-orders in $\lambda_i$'s and leading order in $1/k$ was constructed  in \cite{Georgiou:2017jfi} and will not be needed
for our purposes. It has the remarkable invariance given by
\be
\label{symmetry}
\frak{g}_1\to \frak{g}_2^{-1}\,,\quad
 \frak{g}_2\to \frak{g}_1^{-1}\,,\quad
k_1\to-k_2\,,\quad k_2\to-k_1\,,\quad \lambda_1\to\lambda_1^{-1}\,,\quad \lambda_2\to\lambda_2^{-1}\, ,
\ee
which clearly is not a symmetry of its linearized form \eqn{action}.

Due to the fact that the two terms in the perturbation \eqref{action} have mutually vanishing operator product expansions
there is a factorization of the correlation functions which involve current and bilinear current correlators. In parti\-cular,
the corresponding $\b$-functions take the form of two copies of the $\l$-deformed models \cite{Sagkrioti:2018rwg}.
This construction has been extended to a multi-matrix deformation of an arbitrary number of mutually interacting WZW models
\cite{Georgiou:2017oly}.
Due to this factorization property, it is simpler and equivalent to consider
the single deformed case, $\lambda_2=0$, $\lambda_1=\l$, where the linearized form in $\lambda_{ab}$ 
is also the exact form \cite{Georgiou:2017jfi}
  \begin{equation}
  S_{k_1,k_2}^{\lambda}=S_{k_1}(\mathfrak{g}_1)+S_{k_2}(\mathfrak{g}_2)+
  k\frac{\lambda_{ab}}{\pi}\int \text{d}^2\sigma\, J_{1+}^aJ_{2-}^b\ .
  \label{action1}
  \end{equation}
 For this model the $\beta$-functions have been computed to all-orders in the perturbative $\lambda$-expansion  and up to order  $1/k$ in the large-$k$ expansion in\cite{Sagkrioti:2018rwg}. A slight extension to include diffeomorphisms is worked out in  \autoref{diffsRG} where we refer for details. The end result reads
\be
\label{RGgeneral}
\beta^{ab}=\frac{\text{d}\lambda_{ab}}{\text{d}t}=\frac{1}{2k}\mathcal{N}_{ac}{}^d\left(\mathcal{N}_{bd}{}^{(T)c}+g_{bd}\zeta^c\right)\,,
\ee
with
\ba
\label{skslslss}
&&\mathcal{N}_{ab}{}^c=\mathcal{N}_{ab}{}^c(\l,\lambda_0^{-1})=
(\lambda_{ae}\lambda_{bd}f_{edf}-\lambda_0^{-1}\lambda_{ef}f_{abe})g^{fc}\,,\quad
\mathcal{N}_{ab}{}^{(T)c}=\mathcal{N}_{ab}{}^c(\lambda^T,\lambda_0)\,,\nonumber\\
&& g_{ab}=(\mathbb{I}-\lambda^T\l)_{ab},\quad \tilde{g}_{ab}=(\mathbb{I}-\l\lambda^T)_{ab},\quad
g^{ab}=g_{ab}^{-1},\quad \tilde{g}^{ab}=\tilde{g}_{ab}^{-1}\,,\\
&& \zeta^c=\text{constant}\,,\quad \lambda_0=\sqrt{\frac{k_1}{k_2}}\ .
\nonumber
\ea
The parameter $\lambda_0$ is taken to be less than one with no loss of generality and $\zeta^a$ relates to diffeomorphisms. 
In their absence and for $\l_0=1$ the above were derived in \cite{Sfetsos:2014jfa}.

The structure of this work is the following: In \autoref{Zamolodchikovsec}, we compute the Zamolodchikov's metric in the couplings space
and in \autoref{Weylsec} the exact $C$-function through the
Weyl-anomaly coefficient. As an application, in \autoref{su2example} we present the example of the anisotropic $SU(2)$ case.  In
\autoref{Anomalsec}, we compute the anomalous dimension of the
composite operator $J^a_{1+}J^b_{2-}$ by applying gravitational techniques. Our result for the $C$-function is compatible with the one in
\cite{c-function:2018} for a diagonal and isotropic matrix and has all the
correct properties indicated by Zamolodchikov's $c$-theorem, while the anomalous dimension matrix at the same limit
$\lambda_{ab}=\l\delta_{ab}$ reduces to the one found in \cite{Georgiou:2016zyo}.
Finally, we include two appendices: \autoref{diffsRG} proves the form of the additional (diffeomorphisms) terms in the renormalization
 group (RG) flows of Eq. \eqref{RGgeneral}. In \autoref{Zammetric} we 
derive the general Zamolodchikov metric in the couplings space of the current bilinear operator 
which drives the perturbation away from the UV fixed point.

\section{The exact C-function}
\label{Cfunctionsec}

In this section we compute the $C$-function exactly in { $\left(\lambda_{1,2}\right)_{ab}$} and to leading order in
the large-$k$ expansion.

\subsection{Zamolodchikov's metric}
\label{Zamolodchikovsec}

{  
Following the discussion in \autoref{secintro}, the metric 
takes the form of two copies of the single $\l$-deformed models. Thus, it suffices to 
focus on the special case with $\l_2=0$, $\l_1=\l$ whose effective action is given in \eqref{action1}.}
To proceed, we move to the Euclidean worldsheet with complex coordinates 
$\displaystyle z={1\ov \sqrt{2}} \left(\tau+i\,\sigma\right)$ and $\bar z$, yielding
the action
\begin{equation}
\label{perturbation}
 S_{k_1,k_2}^{\l}=S_{k_1}(\mathfrak{g}_1)+S_{k_2}(\mathfrak{g}_2)
-\frac{\lambda_{ab}}{\pi}\int\text{d}^2z\, {\cal O}_{ab}(z,\bar z)\,,\quad {\cal O}_{ab}(z,\bar z)=J_1^a(z)\bar J_2^b(\bar z)\,,
\end{equation}
where we have rescaled the currents as $J_i^a\to J_i^a/\sqrt{k_i}$, so that they obey
\begin{equation}
J^a_i(z_1)J^b_i(z_2)=\frac{\delta_{ab}}{z_{12}^2}+\frac{if_{abc}}{\sqrt{k_i}}\frac{J_i^c(z_2)}{z_{12}}+\cdots,\quad
z_{12}=z_1-z_2\,,\quad i=1,2
\end{equation}
and accordingly for the anti-holomorphic currents $\bar J_i^a(\bar z)$.

\no
We will need for our purposes the Abelian ($k$-independent) part of the Zamolodchikov metric $G_{ab|cd}$.
 This computation for the perturbation \eqref{perturbation} is performed in detail in \autoref{Zammetric}, where we find the result
\begin{equation}
\label{twopoint}
\langle{\cal O}_{ab}(x_1,\bar x_1){\cal O}_{cd}(x_2,\bar x_2)\rangle_\l
=\frac{G_{ab|cd}}{|x_{12}|^4}\ ,
\end{equation}
where $G_{ab|cd}$ is given by
\begin{equation}
\label{ZamMetric}
G_{ab|cd}=\frac12\left(\tilde g^{-1}\otimes g^{-1}\right)_{ab|cd}=\frac12\tilde g^{ac} g^{bd}\,,
\end{equation}
where $g,\tilde g$ were defined in \eqref{skslslss}.
This is a positive semi-definite matrix since it is the direct product of such matrices.
The inverse metric equals
\begin{equation}
G^{ab|cd}=(G^{-1})_{ab|cd}  =2\left(\tilde g\otimes g\right)_{ab|cd}
=2\tilde g_{ac} g_{bd}\,,\quad G_{ab|mn}G^{mn|cd}=\delta_a^c\delta_b^d\,.
\end{equation}
The corresponding line element in the couplings target space is non-negative
\begin{equation}
\label{Zam.metric}
\text{d}\ell^2=G_{ab|cd}\,\text{d}\lambda_{ab}\,\text{d}\lambda_{cd}\geqslant 0
\end{equation}
and moreover it is invariant under the transformation
 $\l\to\lambda^{-1}$, since
\begin{equation}
g^{-1}\to-\lambda g^{-1}\lambda^T\,,\quad \tilde g^{-1}\to-\lambda^T\tilde g^{-1}\lambda\ .
\end{equation}

\subsection{The Weyl anomaly coefficient}
\label{Weylsec}

In order to compute the $C$-function \eqref{Cfunctiongeneral} for $\s$-models corresponding to \eqref{action} first 
recall its fundamental property
\begin{equation}
\label{sklldldldshsj}
\frac{\text{d}C}{\text{d}t}=\sum\limits_{i=1}^2\beta_i^{ab}\frac{\partial C}{\partial\left(\lambda_i\right)_{ab}}
=24\sum\limits_{i=1}^2G^i_{ab|cd}\beta_i^{ab}\beta_i^{cd}=
12\sum\limits_{i=1}^2\text{Tr}\left(\beta_i^T\tilde g_i^{-1}\beta_i g_i^{-1}\right)\geqslant0\,,
\end{equation}
where we have used \eqref{ZamMetric} and \eqref{Zam.metric}.
The $\beta_i^{ab}$ with $i=1,2$ are the $\b$-functions corresponding
to the two coupling matrices $\left(\lambda_i\right)_{ab}$.
A solution to \eqref{sklldldldshsj} is
\begin{equation}
\label{beta.cfunction}
\left(\beta_i\right)_{ab}=\frac{1}{24}\frac{\partial C}{\partial\left(\lambda_i\right)_{ab}}\,,\quad\text{where:}\quad
\left(\beta_i\right)_{ab}=G^i_{ab|cd}\beta_i^{cd}\,,
\end{equation}
under the assumption that $\left(\beta_i\right)_{ab}\,\text{d}\left(\lambda_i\right)_{ab}$ is a closed one-form.
Integrating \eqref{beta.cfunction} can still be quite laborious and an alternative method needs to be pursued.
We shall demonstrate that for the $\sigma$-model \eqref{action}, the
$C$-function  is given in terms of the Weyl anomaly coefficient \cite{Tseytlin:1987bz,Tseytlin:2006ak}
\begin{equation}
\label{generalC}
\begin{split}
& C_\text{double}=2\,\dim G-3\left(R-\frac{1}{12}H^2+4\nabla^2\Phi-4\left(\del\Phi\right)^2\right)
\\
& \phantom{xxxxx}\ = 2\,\dim G-3\left(R_- + \frac{1}{6}H^2+4\nabla^2\Phi-4\left(\del\Phi\right)^2\right)  
\end{split}
\end{equation}
and that \eqref{beta.cfunction} is indeed solved. In the second line we have used for later convenience
the torsion-full Ricci scalar $\displaystyle R_-=R-\frac14H^2$.

\no
Generically \eqn{generalC} depends explicitly on $X^\mu$ and it is a constant if and only if	
\begin{equation}
\label{sllsdksllsls}
4\frac{\text{d}G_{\mu\nu}}{\text{d}t}\,\partial^\nu\Phi+\frac{\text{d}B_{\nu\rho}}{\text{d}t}H_\mu{}^{\nu\rho}=
2\nabla^\nu\left(\frac{\text{d}G_{\mu\nu}}{\text{d}t}\right)\,,
\end{equation}
where the one-loop $\b$-functions for $G_{\mu\nu}$ and $B_{\mu\nu}$ are given through \cite{honer,Friedan:1980jf,Curtright:1984dz}
\be
\begin{split}
&\frac{\text{d}G_{\mu\nu}}{\text{d}t}=R_{\mu\nu}-\frac14H^2_{\mu\nu}+2\nabla_\mu\partial_\nu\Phi\,,\\
&\frac{\text{d}B_{\mu\nu}}{\text{d}t}=-\frac12\nabla_\rho\left(\text{e}^{-2\Phi}H^\rho{}_{\mu\nu}\right)\,.
\end{split}
\ee
For conformal backgrounds the condition \eqref{sllsdksllsls} is trivially satisfied.

Next, we specialize to the models at hand, whose linearized form was given in \eqref{action}.
Following the discussion in \autoref{secintro}, the
$C$-function takes the form of two copies of the single $\l$-deformed models\footnote{
For a general deformation involving only mutual interactions of the cyclic-type having the form \cite{Georgiou:2017oly}
$\displaystyle \cL_{\rm pert}  =  {k\ov \pi} \sum_{i=1}^n \l^{ab}_{i+1} J_{(i+1)+}^a J_{i-}^b$, with
$ J_{(n+1)\pm}^a= J_{1\pm}^a $,
the expression \eqref{Cfunctiondouble} generalizes to
\begin{equation*}
C_n(\lambda_i,;k_i)=\sum\limits_{i=1}^nC_\text{single}\left(\lambda_i;\sqrt{k_i\,k_{i+1}}\,,\sqrt{\frac{k_i}{k_{i+1}}}\,\right)-(n-1)\,c_\text{UV}\ ,
\quad k_{n+1}=k_1\ , \quad  c_{\rm UV} = \sum_{i=1}^n {2 k_i \dim G\ov 2 k_i + c_G}\ .
\end{equation*}
}
\begin{equation}
\label{Cfunctiondouble}
C_\text{double}(\lambda_1,\lambda_2;k,\lambda_0)=C_\text{single}(\lambda_1;k,\lambda_0)+
C_\text{single}(\lambda_2;k,\lambda_0^{-1})-c_\text{UV}\,,
\end{equation}
where  $C_\text{single}(\l;k,\lambda_0)$, corresponds to the single deformed case
with action \eqn{action1}.
We have chosen the dependence on the levels $k_1$ and $k_2$ via the parameters $0<\l_0<1$ and $k\gg 1$.
The last term in \eqn{Cfunctiondouble} involves the central charge at the UV and has been inserted in order to
 satisfy the conditions
\begin{equation}
C_\text{double}(0,0;k,\lambda_0)=C_\text{single}(0;k,\lambda_0)=C_\text{single}(0;k,\lambda^{-1}_0)=c_\text{UV}\ .
\end{equation}
Explicitly from the standard Sugawara construction
\begin{equation}
c_{\rm UV}= {2 k_1\dim G\ov 2 k_1 + c_G}+ {2 k_2\dim G\ov 2 k_2 + c_G} = 2 \dim G
- {c_G \dim G\ov 2 k} (\lambda_0+\lambda_0^{-1}) + {\cal O}\left(\frac{1}{k^2}\right)\,.
\end{equation}
Hence the computation boils down to determining $C_\text{single}(\l;k,\lambda_0)$.
 This computation heavily depends on several results that can be collectively found in section 2.1.2 of \cite{Sagkrioti:2018rwg}.
Here, the corresponding Weyl anomaly coefficient drastically simplifies since the diffeomorphisms
$\xi_A$ vanish\footnote{
\label{slflldfjdd}
For a reduced $\lambda_{ab}$, an additional
diffeomorphism might be needed for ensuring consistency of the RG flow,
see Eq.\eqref{RGgeneral} and its derivation performed in \autoref{diffsRG}. In that case, the dilaton contribution has
to be included as in Eq. \eqref{generalC}.}
\be
\label{skfldlldkssk}
\xi_A=\om^{-C}{}_{A|C}=0\ ,
\ee
corresponding to a constant dilaton since $\xi_A=2\,\del_A\Phi$.
In this case, the Weyl anomaly coefficient simplifies to
\be
\label{Weylsingle}
C_\text{single}(\l;k,\lambda_0)=2\,\text{dimG}-3\left(R_- + \frac{1}{6}H^2\right)\ .
\ee
We clarify that whereas for $C_\text{double}$ we need to use the action \eqn{action} in its full non-linearity, for
$C_\text{single}$ instead, the simple action \eqn{action1} suffices.

Continuing with our computation, the torsion-full Ricci scaler  $R_-$ can be expressed in terms of the $\b$-function as
\begin{equation}
\label{Riccim}
R_-=-2\,\text{Tr}\left(\frac{\text{d}\l}{\text{d}t}\lambda^T\tilde g^{-1}\right)=\frac{\text{d}}{\text{d}t}\ln\det\tilde g\,.
\end{equation}
Note that this is not invariant under the transformation \eqref{symmetry}
\begin{equation}
R_-\to-2\,\text{Tr}\left(\frac{\text{d}\l}{\text{d}t}\lambda^{-1}\tilde g^{-1}\right)\,.
\end{equation}
Next we evaluate $H^2$, using the components of the three-form $H$ in a convenient frame computed in \cite{Sagkrioti:2018rwg}. We find that
\begin{equation}
\begin{split}
&H^2=\frac{\lambda_0}{k}\left(I_{abc}I_{pqr}\tilde g^{ap}\tilde g^{bq}\tilde g^{cr}+3{\cal N}_{bc}{}^d{\cal N}_{qr}{}^eg^2_{de}\tilde g^{bq}\tilde g^{cr}
+c_G\,\text{dim}G\right)\,,\\
&I_{abc}=\lambda_0^{-1}f_{abd}\,\tilde g_{cd}+{\cal N}_{bc}{}^d\left(\lambda^T\tilde g\right)_{da}+{\cal N}_{ca}{}^d\left(\lambda^T\tilde g\right)_{db}\ ,
\end{split}
\end{equation}
which similarly to $R_-$ is not invariant under the transformation \eqref{symmetry}.
Then, plugging the above into \eqref{Weylsingle} we find after certain algebraic manipulations that
\begin{equation}
\begin{split}
C_\text{single}(\l;k,\lambda_0)&=\left(2-\frac{c_G\lambda_0}{2k}\right)\text{dimG}+6\,\text{Tr}\left(\b\lambda^T\tilde{g}^{-1}\right)\\
&-\frac{\lambda_0}{2k}\bigg(I_{abc}I_{pqr}\tilde g^{ap}\tilde g^{bq}\tilde g^{cr}
+3{\cal N}_{bc}{}^d{\cal N}_{qr}{}^eg^2_{de}\tilde g^{bq}\tilde g^{cr}\bigg)\,.
\label{complete Csingle}
\end{split}
\end{equation}
Finally, we should substitute the above into \eqn{Cfunctiondouble} and verify,
using \eqref{RGgeneral} and \eqref{ZamMetric},
 that the system of differential equations \eqn{beta.cfunction} is indeed obeyed
  without any diffeomorphisms.
 This is a formidable task which we did not complete in full generality.
We have checked with Mathematica in various examples, involving the groups $SU(2),SU(3),SP(4),G_2$ and for various
couplings $\left(\lambda_i\right)_{ab}$, that indeed this is the case. 
This leaves little doubt that, with the above data, \eqn{beta.cfunction}
is obeyed in general.

\no
For an isotropic coupling $\lambda_{ab}=\l\delta_{ab}$, \eqref{complete Csingle} reduces to Eq. (2.14) of \cite{c-function:2018}, corresponding to a flow from $G_{k_1}\times G_{k_2}$ in
the UV point ($\lambda=0$) to $G_{k_1}\times G_{k_2-k_1}$ in the IR point ($\lambda=\lambda_0$) \cite{Georgiou:2017jfi}.
 For isotropic couplings $\left(\lambda_{1,2}\right)_{ab}=\lambda_{1,2}\,\delta_{ab}$,
\eqref{Cfunctiondouble} reduces to Eq. (2.11) of \cite{c-function:2018}, corresponding to a flow from
$G_{k_1}\times G_{k_2}$ in the UV point ($\lambda_{1,2}=0$)
to $\frac{G_{k_1}\times G_{k_2-k_1}}{G_{k_2}}\times G_{k_2-k_1}$ in the IR point ($\lambda_{1,2}=\lambda_0$) \cite{Georgiou:2017jfi}.

\no
Last but not least, $C_\text{single}$ is invariant under the transformation \eqref{symmetry}, up to a constant
\begin{equation}
\label{skfosoofkd}
C_\text{single}(\lambda^{-1};-k,\lambda_0^{-1})-C_\text{single}(\l;k,\lambda_0)=
\frac{c_G\dim G}{2k}\left(\lambda_0+\lambda_0^{-1}\right)\,.
\end{equation}
Subsequently, one can use the above and \eqref{Cfunctiondouble} to prove that $C_\text{double}(\lambda_1,\lambda_2;k,\lambda_0)$,
is invariant under the non-perturbative symmetry transformation \eqref{symmetry}.
Interestingly, equality of the $C$-functions under this transformation is achieved only when both couplings 
are allowed to change under the RG flow so that they both may reach their common fixed value in the IR.

\subsection{The anisotropic $SU(2)$ example}
\label{su2example}

In the anisotropic $SU(2)$ case we have six couplings, parameterized as
\begin{equation}
\label{jfufisodhjd}
\left(\lambda_{1}\right)_{ab}=\diag\left(\lambda_1,\lambda_2,\lambda_3\right)\,,
\quad\text{and}\quad\left(\lambda_{2}\right)_{ab}=\diag\left(\tilde\lambda_1,\tilde\lambda_2,\tilde\lambda_3\right)\,,
\end{equation}
with the metrics of the composite operator given by
\begin{equation}
\label{metricsu2}
G_{ab}=\frac{\delta_{ab}}{2\left(1-\lambda_a^2\right)^2}\,,\quad
\widetilde G_{ab}=\frac{\delta_{ab}}{2\left(1-\tilde\lambda_a^2\right)^2}\,,
{ \quad a=1,2,3.}
\end{equation}
To compute the exact in $\l$'s and leading order in $k$, $\b$-functions of this model, we employ
the results of \cite{LeClair:2001yp,Sagkrioti:2018rwg}. We find that \cite{Sfetsos:2014jfa}
\begin{equation}
\label{betasu2}
\b^1=\frac{\text{d}\lambda_1}{\text{d}t}=-\frac{2\left(1+\lambda_1^2\right)\lambda_2\lambda_3-\left(\lambda_0+\lambda_0^{-1}\right)\lambda_1\left(\lambda_2^2+\lambda_3^2\right)}
{k\left(1-\lambda_2^2\right)\left(1-\lambda_3^2\right)}
\end{equation}
and cyclic in $\lambda_{1,2,3}$. The $\b$-functions for the $\tilde\lambda_a$, are obtained by simply
relabeling $\lambda_a\to\tilde\lambda_a$.
The fixed points of the $\beta$-functions and the corresponding  CFTs read \cite{Georgiou:2017jfi}
\begin{equation}
\label{fixedsu2}
\begin{split}
&\text{UV}:\quad \lambda_{1,2}=0=\tilde\lambda_{1,2}\,,\quad SU(2)_{k_1}\times SU(2)_{k_2}\,,\\
&\text{IR}_1:\quad  \lambda_a=\lambda_0=
\tilde\lambda_a\,,\quad \frac{SU(2)_{k_1}\times SU(2)_{k_2-k_1}}{SU(2)_{k_2}}\times SU(2)_{k_2-k_1}\,,\\
&\text{IR}_2:\quad \lambda_a=\lambda_0\,,\quad\tilde\lambda_a=0\,,\quad SU(2)_{k_1}\times SU(2)_{k_2-k_1}\,.
\end{split}
\end{equation}
A comment is in order related to the UV fixed point.
At first the choice of $\lambda_{1,2}=0=\tilde\lambda_{1,2}$ is just a matter of convention as
other pairs of $\lambda$'s could have been chosen.
One can show that this point corresponds to an exact CFT, as
the parameters $(\lambda_3,\tilde\lambda_3)$ can be absorbed by an $O(4,4)$ duality transformation on the
exact $SU(2)_{k_1}\times SU(2)_{k_2}$ string background. This is consistent with the perturbation being $\displaystyle \cL_{\rm pert}= {k\ov \pi} \left(\l_3 J_{1+}^3 J_{2-}^3 + \tilde \l_3 J_{2+}^3 J_{1-}^3 \right)$,
i.e. in the Cartan subalgebra of $SU(2)\times SU(2)$, and hence exactly marginal.
\begin{comment}
\ffotnote{In the standard
notation the $O(4,4)$-transformation is obtained by exponentiating the generator
\be
\left(
   \begin{array}{cc}
     A & B \\
     C & -A^T \\
   \end{array}
 \right)\ , A = \ha {\rm skew-diagonal}(\l_0 \l_3,\l_0 \tilde \l_3,\l_0^{-1} \tilde \l_3,\l_0^{-1}\l_3)\ ,\qq B = \ha {\rm skew-diagonal}(\l_3,-\tilde \l_3,\tilde \l_3,-\l_3)\ ,\quad C=-B\ .
\ee
\end{comment}

Defining $\beta_a=G_{ab}\beta^b$, one can prove that $\beta_a\text{d}\lambda_a$ is a closed one-form and
similarly to \eqref{beta.cfunction} we find that
\begin{equation}
\label{bisu2}
\beta_a=\frac{1}{24}\frac{\partial C_\text{double}}{\partial\lambda_a}\,,\quad \tilde\beta_a=
\frac{1}{24}\frac{\partial C_\text{double}}{\partial\tilde\lambda_a}\,,
\end{equation}
with
\begin{eqnarray}
\label{cfunctionsu2}
&&{ C_\text{double}(\lambda_a,\tilde\lambda_a;k,\lambda_0)=c_\text{UV}-\frac{6}{k}\left(f(\lambda_a;\lambda_0)+f(\tilde\lambda_a;\lambda_0)\right)\,,}\\
&&f(\lambda_a;\lambda_0)=\frac{4\lambda_1\lambda_2\lambda_3-\left(\lambda_0+\lambda_0^{-1}\right)\left(\lambda_1^2\lambda_2^2+\lambda_2^2\lambda_3^2+\lambda_1^2\lambda_3^2-\lambda_1^2\lambda_2^2\lambda_3^2\right)}
{(1-\lambda_1^2)(1-\lambda_2^2)(1-\lambda_3^2)}\,,\nonumber
\end{eqnarray}
where $c_\text{UV}$ is the central charge at the UV fixed point \eqref{fixedsu2}, namely: $\lambda_{1,2}=0=\tilde\lambda_{1,2}$
\begin{equation}
\label{centralUV}
c_\text{UV}=6-\frac6k\left(\lambda_0+\lambda_0^{-1}\right)+{\cal O}\left(\frac{1}{k^2}\right)\,.
\end{equation}
Before closing this section note that the $C$-function \eqref{cfunctionsu2} is invariant under
the transformation
\begin{equation}
\lambda_a\to\lambda_a^{-1}\,,\quad \tilde\lambda_a\to\tilde\lambda_a^{-1}\,,\quad k\to-k\,,\quad \lambda_0\to\lambda_0^{-1}\,.
\end{equation}
and it reproduces the central charges at the UV and the $\text{IR}_{1,2}$ fixed points \eqref{fixedsu2}.

\section{Anomalous dimension of the bilinear current}
\label{Anomalsec}

In this section we compute the anomalous dimension matrix for the bilinear current operator.
{  To do so, we recall results of \cite{Kutasov:1989dt}
\be
\langle{\cal O}_{ab}(x_1,\bar x_1){\cal O}_{cd}(x_2,\bar x_2)\rangle_{\l,k}
=\frac{1}{|x_{12}|^4}\left(G_{ab|cd}+\g_{ab|cd}\ln\frac{\varepsilon^2}{|x_{12}|^2}\right)\,,
\ee
where
\begin{equation}
\label{general.anomalous}
\gamma_{ab}{}^{cd}=\nabla_{ab}\b^{cd}+\nabla^{cd}\beta_{ab}=\nabla_{ab}\b^{cd}+G_{ab|mn}G^{cd|pq}\nabla_{pq}\b^{mn}\,,
\end{equation}
with $\nabla_{ab}\b^{cd}=\partial_{ab}\b^{cd}+\G^{cd}_{ab|mn}\b^{mn}$. The $\G^{cd}_{ab|mn}$ are the standard Christoffel symbols 
and can be computed throughout the Zamolodchikov metric  \eqn{ZamMetric} 
\be
\Gamma^{p_1p_2}_{m_1m_2|n_1n_2}=\frac{1}{2}G^{p_1p_2|q_1q_2}\bigg(\partial_{m_1m_2}G_{q_1q_2|n_1n_2} +\partial_{n_1n_2}G_{q_1q_2|m_1m_2}-\partial_{q_1q_2}G_{m_1m_2|n_1n_2} \bigg)\ ,
\ee
where we denoted $\displaystyle \partial_{m_1m_2}=\frac{\partial}{\partial\lambda_{m_1m_2}}$.
With the help of the identity $ \l g^{-1}=\tilde{g}^{-1}\l$, the result can be brought into the form
\be
\Gamma^{p_1p_2}_{m_1m_2|n_1n_2}=\delta^{p_1}_{n_1}\delta^{p_2}_{m_2}\big(\l g^{-1}\big)_{m_1n_2}+
\delta^{p_1}_{m_1}\delta^{p_2}_{n_2}\big(\l g^{-1}\big)_{n_1m_2}\ .
\ee
After some algebra we find the anomalous dimension matrix \eqref{general.anomalous}}
\ba
&& \gamma_{ab}{}^{cd}=\frac{1}{2k}\bigg( \delta_{ap}\delta_{bq}\delta^{cm}\delta^{dn}+
G_{ab|mn}G^{cd|pq}  \bigg)\bigg\{ \mathcal{N}^{(T)k}_{ni}\bigg[\big(\delta_{mp}\lambda_{ks}-\delta_{pk}\lambda_{ms} \big)f_{qsf}g^{fi}- \lambda_0^{-1}f_{mkp}g^{qi}
\nonumber
\\
&& +\lambda_{pl}\big(\mathcal{N}_{mk}{}^lg^{qi}+\mathcal{N}_{mk}{}^qg^{li} \big)  \bigg]+
\mathcal{N}_{mk}{}^i\bigg[ \big(\delta_{nq}\lambda_{si}-\delta_{iq}\lambda_{sn}  \big)f_{psf}\tilde{g}^{fk}- \lambda_0f_{niq}\tilde{g}^{pk}
\\
&& +\lambda_{lq}\big(\mathcal{N}_{ni}^{(T)p}\tilde{g}^{lk}+\mathcal{N}_{ni}^{(T)l}\tilde{g}^{pk} \big)   \bigg]
+\delta^m_p\mathcal{N}_{lr}{}^s\mathcal{N}^{(T)r}_{ns}\big(\lambda g^{-1} \big)_{lq}
+\delta^n_q\mathcal{N}_{mr}{}^s\mathcal{N}^{(T)r}_{ls}\big(\l g^{-1}\big)_{pl}\bigg\}\,.
\label{anomalous dim}
\nonumber
\ea
This expression is quite involved and we could not further simplify it. Still, 
it  transforms as a mixed tensor under the duality-type symmetry
\begin{equation}
\gamma_{ab}{}^{cd}\to \lambda_{ea}\lambda_{bf}\lambda^{-1}_{cg}\lambda^{-1}_{hd}\gamma_{ef}{}^{gh},
\end{equation}
as expected.
Specializing to an isotropic coupling $\lambda_{ab}=\l\delta_{ab}$, we obtain
\begin{align}
\begin{split}
\gamma_{ab}{}^{cd}=&c_G\lambda^2\frac{(1+\lambda^2)(\lambda_0+\lambda_0^{-1})-4\l}{(1-\lambda^2)^3}\delta_{ac}\delta_{bd}
+\frac{\lambda^2(\lambda_0+\lambda_0^{-1})-2\l}{(1-\lambda^2)^2}f_{ace}f_{bde}\\
&+\lambda^2\frac{(1+3\lambda^2)(\lambda_0+\lambda_0^{-1})-2\l(3+\lambda^2)}{(1-\lambda^2)^3}f_{ade}f_{bce}\,.
\end{split}
\end{align}
The corresponding anomalous dimension  is found from the eigenvalue problem
\be
\gamma_{ab}{}^{cd}\delta_{cd}=\gamma\,\delta_{ab}\,,
\ee
which coincides with that in Eq. (2.16) of \cite{Georgiou:2016zyo}
\begin{equation}
\gamma=c_G\,\l\frac{3(\lambda_0+\lambda_0^{-1})\l(1+\lambda^2)-2(1+4\lambda^2+\lambda^4)}{k(1-\lambda^2)^3}\,.
\end{equation}
Other checks for equal level include the $SU(2)$ case with anisotropic coupling and the
two coupling case using a symmetric coset, see Eq.(3.11) and the equation after (3.15) of
\cite{Georgiou:2015nka}, respectively. Finally, we note  that when 
the current bilinear is restricted to the Cartan subgroup then  \eqn{general.anomalous} for the corresponding anomalous dimension  vanishes, in accordance with the fact that the perturbation is then exactly marginal. 

\section{Outlook}

In this paper we presented the exact $C$-function for the doubly $\l$-deformed models for generic couplings. This was done by computing the general metric in the space of couplings and subsequently,
incorporating the exact $\beta$-function for these models. 
We demonstrated that the Weyl anomaly is indeed  Zamolodchikov's $C$-function. 
In addition, we have computed the anomalous dimension matrix of the composite current bilinear operator driving the perturbation away from conformality.

Our results also provide $C$-functions for the so-called $\eta$-deformations for group and coset spaces introduced in \cite{Klimcik:2002zj,Klimcik:2008eq,Klimcik:2014,Delduc:2013fga,Delduc:2013qra}. The reason is that these models are related to 
symmetric $\l$-deformations ({ that is when the levels of the CFTs are equal}) via Poisson--Lie T-duality and appropriate analytic continuations  \cite{Vicedo:2015pna,Hoare:2015gda,Sfetsos:2015nya,Klimcik:2015gba}. { In particular, the background fields, the $\beta$-functions, the $C$-functions, etc map to each other. However, the analytic transformation spoils the UV behavior of the $\eta$-deformed models, as compared to that of the $\lambda$-models. In particular, there is no UV fixed point and they generically possess cyclic RG-flows \cite{Appadu:2017bnv}. }

Finally, we note that the all loop effective action representing, for small couplings, simultaneously self and mutually interacting current algebra CFTs realized by two different WZW models were constructed in \cite{Georgiou:2018hpd}.
It will be very interesting to extend the results of the present paper in this most general case as well.

\section*{Acknowledgments}

K. Sfetsos would like to thank the Theoretical Physics Department of CERN for hospitality and
financial support during part of this research.

\appendix

\section{Renormalization and diffeomorphisms}
\label{diffsRG}

The scope of this appendix is to work out the presence of diffeomorphisms $\xi$'s for the RG flows \eqref{RGgeneral},
of the $\sigma$-model \eqref{action1}, which were explicitly worked out in \cite{Sagkrioti:2018rwg}.
Consider the generic one-loop RG flow \cite{honer,Friedan:1980jf,Curtright:1984dz}
\be
\frac{\text{d}}{\text{d}t}\left(G_{MN}+B_{MN}\right)=R^-_{MN}+\nabla^+_N\xi_M+\nabla_{[M}\zeta_{N]}\,,\quad t=\ln\mu^2\,,
\ee
where $\mu$ is the RG scale, $R^-_{MN}$ is the torsion-full Ricci and $(\xi_M,\zeta_M)$ correspond to diffeomorphisms
and gauge transformations respectively. For the scope of this appendix, it suffices to only consider $\xi_M$.
The above expression can be rewritten equivalently
in the tangent frame $\text{e}^A=\text{e}^A{}_M\,\text{d}X^M$
\be
\label{kfkdllsjsks}
\frac{\text{d}}{\text{d}t}\left(G_{AB}+B_{AB}\right)=R^-_{AB}+\nabla^-_B\xi_A\,.
\ee
The term $\xi_A$ involves two contributions \cite{Sagkrioti:2018rwg}
\be
\xi_A=\om^{-C}{}_{A|C}+\hat\xi_A\,,
\ee
where the first one is vanishing through \eqref{skfldlldkssk} and the second one incorporates additional
diffeomorphisms that might be needed for ensuring consistency of the
RG flow in cases with a reduced $\lambda_{ab}$. Next we rewrite the $\hat\xi_A$ term
\be
\label{skflldldl}
\nabla_B^-\hat\xi_A=\text{e}_B{}^M\left(\del_M\hat\xi_A+\om^-_A{}^C{}_{|M}\,\hat\xi_C\right)\,,
\ee
where $\text{e}_A{}^M$ is the inverse of $\text{e}^A{}_M$, i.e. $\text{e}^A{}_M\,\text{e}_A{}^N=\delta_M^N$.
Plugging \eqref{skflldldl} into \eqref{kfkdllsjsks} along with the results of section 2.1.2 of \cite{Sagkrioti:2018rwg}, leads to 
the consistent set of RG flows
\be
\frac{\text{d}\lambda_{ab}}{\text{d}t}=\frac{1}{2k}\mathcal{N}_{ac}{}^d\left(\mathcal{N}_{bd}{}^{(T)c}-
\lambda_0g_{bd}\tilde g^{ce}\hat\xi_e\right)\,,\quad
\hat\xi_e=\text{constant}\,,
\ee
which take the form of  \eqref{RGgeneral} where $\zeta^c=-\lambda_0\tilde g^{ce}\hat\xi_e$.

\section{Computation of Zamolodchikov's metric}
\label{Zammetric}

{  In this appendix we compute} the Zamolodchikov metric \eqref{ZamMetric} for the composite operator ${\cal O}_{ab}$
in \eqref{perturbation}.
The metric in the couplings space can be found through the two-point function  \cite{Zamolo:1986}, given in \eqref{twopoint}. Following the lines of appendix A.2 in \cite{Georgiou:2015nka}, we can write the two-point function as a series
expansion
\begin{equation}
\begin{split}
G_{ab|cd}&=|x_{12}|^4\langle{\cal O}_{ab}(x_1,\bar x_1){\cal O}_{cd}(x_2,\bar x_2)\rangle_\l\\
&=G^{(0)}_{ab|cd}+\sum_{n=1}^{\infty}\left(\prod_{i=1}^{2n}\lambda_{a_ib_i}\right)\frac{G^{(2n)}_{aa_1\cdots a_{2n}c|bb_1\cdots b_{2n}d}}{\pi^{2n}(2n)!}\ ,
\label{series}
\end{split}
\end{equation}
where
\begin{equation}
\label{sllldfkdkd}
\begin{split}
& \frac{G^{(2n)}_{aa_1...a_{2n}c|bb_1\cdots b_{2n}d}}{|x_{12}|^4}=\int\text{d}^2z_{1\dots 2n} \braket{J_1^a(x_1)J_1^{a_1}(z_1)\cdots J^{a_{2n}}_1(z_{2n})J^c_1(x_2)}
\\
&\qq\qq\qq\qq\qq\qq \braket{\J_2^b(\bar{x}_1)\J^{b_1}_2(\z_1)\cdots \J^{b_{2n}}_2(\z_{2n})\J^d_2(\bar{x}_2)}\ ,
\end{split}
\end{equation}
with the two-point function of ${\cal O}_{ab}$ evaluated at the conformal point
\begin{equation}
G_{ab|cd}^{(0)}=|x_{12}|^4\langle{\cal O}_{ab}(x_1,\bar x_1){\cal O}_{cd}(x_2,\bar x_2)\rangle_\text{CFT}=\delta_{ac}\delta_{bd}=(\mathbb{I}\otimes\mathbb{I})_{ab|cd}\,.
\label{zero}
\end{equation}

Next we work out \eqref{sllldfkdkd} by performing the appropriate contractions avoiding bubble 
and disconnected diagrams and keeping only the Abelian part, 
we find the recursive relation
\begin{align}
\begin{split}
\frac{1}{\pi^2}G^{(2n)}_{aa_1\cdots a_{2n}c|bb_1\cdots b_{2n}d}&=(2n-1)(2n-2)\delta_{a_1a_2}\delta_{b_1b_3}G^{(2n-2)}_{aa_3a_4\cdots a_{2n}c|bb_2b_4\cdots b_{2n}d}\\
& +2(2n-1)\delta_{aa_1}\delta_{b_1b_2}G^{(2n-2)}_{a_2a_3..a_{2n}c|bb_3b_4\cdots b_{2n}d}\\
&  +2(2n-1)\delta_{a_1a_2}\delta_{bb_1}G^{(2n-2)}_{aa_3a_4\cdots a_{2n}c|b_2b_3\cdots b_{2n}d}\,.
\label{recursion}
\end{split}
\end{align}
This is solved by
\begin{align}
\begin{split}
\left(\prod_{i=1}^{2n}\lambda_{a_ib_i}\right)G^{(2n)}_{aa_1\cdots a_{2n}c|bb_1\cdots b_{2n}d}= 
\pi^{2n}(2n)!\sum_{m=0}^{n}(\lambda\lambda^T)^m_{ac}(\lambda^T\lambda)^{n-m}_{bd},\quad n\geqslant 1\ ,
\label{solution}
\end{split}
\end{align}
a fact that can be proven by induction as follows:
\begin{itemize}
\item It is obvious that for $n=1$ \eqref{solution} holds, since from \eqref{recursion} and (\ref{zero}) we have that
\begin{align*}
\lambda_{a_1b_1}\lambda_{a_2b_2}G^{(2)}_{aa_1a_2c|bb_1b_2d}&=2\pi^2\big(\lambda_{a_1b_1}\lambda_{a_2b_2}\delta_{aa_1}\delta_{b_1b_2}G^{(0)}_{a_2c|bd}+\lambda_{a_1b_1}\lambda_{a_2b_2}\delta_{a_1a_2}\delta_{bb_1}G^{(0)}_{ac|b_2d} \big)\\
&= 2\pi^2\big((\lambda\lambda^T)_{ac}\delta_{bd}+\delta_{ac} (\lambda^T\lambda)_{bd} \big)\,.
\end{align*}
\item 
We assume that \eqref{solution} holds for any order up to $n-1$
\begin{align}
\begin{split}
\left(\prod_{i=1}^{2n-2}\lambda_{a_ib_i}\right)G^{(2n-2)}_{aa_1\cdots a_{2n-2}c|bb_1\cdots b_{2n-2}d}=
 \pi^{2n-2}(2n-2)!\sum_{m=0}^{n-1}(\lambda\lambda^T)_{ac}^m(\lambda^T\lambda)_{bd}^{n-m-1}\,.
 \label{n-1}
\end{split}
\end{align}
\item We prove that \eqref{solution} holds for n.
By multiplying \eqref{recursion} with the $\l$'s we find
\begin{align*}
\frac{1}{\pi^2}&\left(\prod_{i=1}^{2n}\lambda_{a_ib_i}\right)G^{(2n)}_{aa_1\cdots a_{2n}c|bb_1\cdots b_{2n}d}=\\
&=(2n-1)(2n-2)(\lambda\lambda^T\l)_{a_3b_2}\left(\prod_{i=4}^{2n}\lambda_{a_ib_i}\right)
G^{(2n-2)}_{aa_3a_4\cdots a_{2n}c|bb_2b_4\cdots b_{2n}d}\\
&+2(2n-1)(\lambda\lambda^T)_{aa_2}
\left(\prod_{i=3}^{2n}\lambda_{a_ib_i}\right)G^{(2n-2)}_{a_2a_3\cdots a_{2n}c|bb_3b_4\cdots b_{2n}d} \\
&+2(2n-1)(\lambda^T\lambda)_{bb_1}
\left(\prod_{i=3}^{2n}\lambda_{a_ib_i}\right)G^{(2n-2)}_{aa_3\cdots a_{2n}c|b_1b_3b_4\cdots b_{2n}d}\,.
\end{align*}
For the last two terms we can easily substitute \eqref{n-1} for $G^{(2n-2)}$. However, since the contracted indices of the first term do not follow the pattern of 	\eqref{n-1}, a bit more work in needed.
In the first line, we substitute $G^{(2n-2)}$ by its recursive relation \eqref{recursion}. We have
\begin{align*}
\frac{1}{\pi^4}&\left(\prod_{i=1}^{2n}\lambda_{a_ib_i}\right)G^{(2n)}_{aa_1\cdots a_{2n}c|bb_1\cdots b_{2n}d}=
\\
&=(2n-1)\cdots (2n-4)(\lambda\lambda^T\lambda)_{a_3b_2}\delta_{a_3a_4}\delta_{b_2b_5}\left(\prod_{i=4}^{2n}\lambda_{a_ib_i}\right)
G^{(2n-4)}_{aa_5a_6\cdots a_{2n}c|bb_4b_6\cdots b_{2n}d}\\
&+2(2n-1)\cdots (2n-3)(\l\lambda^T\lambda)_{a_3b_2}\delta_{aa_3}\delta_{b_2b_4}	\left(\prod_{i=4}^{2n}\lambda_{a_ib_i}\right)
G^{(2n-4)}_{a_4a_5\dots a_{2n}c|bb_5b_6\dots b_{2n}d}\\
&+2(2n-1)\cdots (2n-3)(\l\lambda^T\lambda)_{a_3b_2}\delta_{a_3a_4}\delta_{bb_2}
\left(\prod_{i=4}^{2n}\lambda_{a_ib_i}\right)G^{(2n-4)}_{aa_5\cdots a_{2n}c|b_4b_5\cdots b_{2n}d}\\
&+ 2\pi^{2n-4}(2n-1)!(\lambda\lambda^T)_{aa_2}\sum_{m=0}^{n-1}(\lambda\lambda^T)^m_{a_2c}(\lambda^T\lambda)^{n-m-1}_{bd}\\
&+ 2\pi^{2n-4}(2n-1)!(\lambda^T\lambda)_{bb_1}\sum_{m=0}^{n-1}(\lambda\lambda^T)^m_{ac}(\lambda^T\lambda)^{n-m-1}_{b_1d}\,,
\end{align*}
where in the second and third line we can use \eqref{n-1} for $G^{(2n-4)}$. We end up with
\begin{align*}
\frac{1}{\pi^4}&\left(	\prod_{i=1}^{2n}\lambda_{a_ib_i}\right)G^{(2n)}_{aa_1\cdots a_{2n}c|bb_1\cdots b_{2n}d}=\\
&=(2n-1)\cdots (2n-4)(\lambda^T\lambda\lambda^T\lambda\lambda^T)_{b_4a_5}\left(\prod_{i=6}^{2n} \lambda_{a_ib_i}\right)
G^{(2n-4)}_{aa_5a_6\cdots a_{2n}c|bb_4b_6\cdots b_{2n}d} \\
&+ 2\pi^{2n-4}(2n-1)!(\lambda\lambda^T\l\lambda^T)_{aa_4}
\sum_{m=0}^{n-2}(\lambda\lambda^T)^m_{a_4c}(\lambda^T\lambda)^{n-m-2}_{bd}\\
&+ 2\pi^{2n-4}(2n-1)!(\lambda^T\l\lambda^T\lambda)_{bb_4}
\sum_{m=0}^{n-2}(\lambda\lambda^T)^m_{ac}(\lambda^T\lambda)^{n-m-2}_{b_4d} \\
&+ 2\pi^{2n-4}(2n-1)!(\lambda\lambda^T)_{aa_2}\sum_{m=0}^{n-1}(\lambda\lambda^T)^m_{a_2c}(\lambda^T\lambda)^{n-m-1}_{bd}\\
&+ 2\pi^{2n-4}(2n-1)!(\lambda^T\lambda)_{bb_1}\sum_{m=0}^{n-1}(\lambda\lambda^T)^m_{ac}(\lambda^T\lambda)^{n-m-1}_{b_1d}\,.
\end{align*}
By continuing the recursion of the first line down to $G^{(0)}$ (where the term containing internal-internal contractions has a vanishing coefficient) only terms of internal-external contractions survive, giving
\begin{equation*}
\begin{split}
&\left(\prod_{i=1}^{2n}\lambda_{a_ib_i}\right)G^{(2n)}_{aa_1\cdots a_{2n}c|bb_1\cdots b_{2n}d}=
\\
&\qq = 2(2n-1)!\pi^{2n}
\sum_{p=1}^{n}\sum_{m=0}^{n-p}\left((\lambda\lambda^T)^{m+p}_{ac}(\lambda^T\lambda)_{bd}^{n-m-p}+
(\lambda\lambda^T)_{ac}^m(\lambda^T\lambda)_{bd}^{n-m}\right)\,,
\end{split}
\end{equation*}
where the double sum rewrites to
\begin{equation*}
\sum_{p=1}^{n}\sum_{m=0}^{n-p}\left((\lambda\lambda^T)^{m+p}_{ac}(\lambda^T\lambda)_{bd}^{n-m-p}+
(\lambda\lambda^T)_{ac}^m(\lambda^T\lambda)_{bd}^{n-m}  \right)=
n\sum_{m=0}^{n}(\lambda\lambda^T)_{ac}^m(\lambda^T\lambda)_{bd}^{n-m}\,.
\end{equation*}	
Assembling all these together we obtain
\be
\left(\prod_{i=1}^{2n}\lambda_{a_ib_i}\right)G^{(2n)}_{aa_1\cdots a_{2n}c|bb_1\cdots b_{2n}d}
= \pi^{2n}(2n)!\sum_{m=0}^{n}(\lambda\lambda^T)_{ac}^m(\lambda^T\lambda)_{bd}^{n-m}\,.
\ee
\end{itemize}
Using the latter into \eqref{series} we have
\begin{align}
\label{kflsldhfjdkdl}
\begin{split}
G_{ab|cd}&= \sum_{n=0}^{\infty}\sum_{m=0}^{n}(\lambda\lambda^T)^m_{ac}(\lambda^T\lambda)^{n-m}_{bd}=
 \sum_{n=0}^{\infty}(\lambda\lambda^T)^n_{ac}\times
\sum_{m=0}^{\infty}(\lambda^T\lambda)^m_{bd}\\
&= (\mathbb{I}-\lambda\lambda^T)_{ac}^{-1}(\mathbb{I}-\lambda^T\lambda)_{bd}^{-1}= \tilde{g}^{ac}g^{bd}
= \big(\tilde{g}^{-1}\otimes g^{-1} \big)_{ab|cd}\,.
\end{split}
\end{align}

A comment is in order related to the additional scaling factor $1/2$ in \eqref{ZamMetric} versus \eqref{kflsldhfjdkdl} 
which contains no such factor.
To understand its appearance we consider the doubled deformed action \eqref{action} with  $\l_1=\l_2=\l$.
Analytically continuing to a Euclidean worldsheet and rescaling the currents as $J_i^a\to J_i^a/\sqrt{k_i}$ 
(as in Eq. \eqref{perturbation}), we obtain
\begin{equation}
\begin{split}
&S_{k_1,k_2}^{\l}=S_{k_1}(\mathfrak{g}_1)+S_{k_2}(\mathfrak{g}_2)
-\frac{\l_{ab}}{\pi}\int\text{d}^2z\,\widehat{\cal O}_{ab}(z,\bar z)+\cdots\,,\\ 
&\widehat{\cal O}_{ab}={\cal O}_{ab}+ \widetilde{\cal O}_{ab}\,,\quad
{\cal O}_{ab}=J_1^a(z)\bar J_2^b(\bar z)\,,\quad \widetilde{\cal O}_{ab}=J_2^a(z)\bar J_1^b(\bar z)\,.
\end{split}
\end{equation}
We are going to normalize the two-point function of $\widehat{\cal O}_{ab}$ to one, so that it matches with 
the conventions used in the proof of the $c$-theorem \eqref{Cfunctiongeneral} in \cite{Zamolo:1986}.
This normalization introduces the additional scaling factor $1/2$ in \eqref{ZamMetric}. 
Note that for the single $\l$-deformed model \cite{Sfetsos:2013wia},
the analogue scaling factor is one \cite{Kutasov:1989dt,Georgiou:2015nka}.

%\section*{Acknoledgments}
	
%%%%%%%%%%%%%%%%%%%%%%%%%%%%%%%%%%%%%%%%%%%%%%%%%%%%%%

\end{document}